\documentclass[11pt]{article}
\pdfoutput=1 
\usepackage{nicefrac} 
\usepackage{jheppubg2} 
\notoc 
\usepackage{accents} 
\usepackage{comment} 
\usepackage[cmyk]{xcolor} 
\usepackage{tikz} 
\usepackage{tikz-3dplot}
\usetikzlibrary{shapes,arrows,decorations.markings} 
\pagecolor{white} 

\definecolor{lyell}{cmyk}{0,0,1,.05}
\definecolor{dyell}{cmyk}{0,0,1,.2}
\definecolor{lgree}{cmyk}{.8, 0, .8, 0.08}
\definecolor{lblue}{cmyk}{.8, .4, 0, .1}
\definecolor{black}{cmyk}{0,0,0,1}
\definecolor{white}{cmyk}{0,0,0,.01}
\definecolor{lgray}{cmyk}{0,0,0,.15}
\definecolor{mgray}{cmyk}{0,0,0,.5}
\definecolor{myell}{cmyk}{0,0,1,.07}
\definecolor{mrora}{cmyk}{0,.5,.9,.02}
\definecolor{mred}{cmyk}{0,.7,.7,0.15}
\definecolor{mygre}{cmyk}{.3,0,.75,.035}
\definecolor{mgree}{cmyk}{.9,0,.9,0.4}
\definecolor{mbvio}{cmyk}{.3,.4,0,0..03}
\definecolor{mblue}{cmyk}{.7,.35,0,0.2}

\def \pscale{1.5}

\def \gplu#1{\tikz\draw[scale=\pscale, color=mgray, fill={#1}, line width=0.1ex, rounded corners=0.05ex, line cap=round] (0,0) -- (0.25ex,0.43ex) -- (0.25ex,0.87ex) -- (0.5ex,0.87ex) -- (1ex,0) -- (0,0);}
\def \gpld#1{\tikz\draw[scale=\pscale, color=mgray, fill={#1}, line width=0.1ex, rounded corners=0.05ex, line cap=round] (0,0) -- (0,0.43ex) -- (0.25ex,0.43ex) -- (0.5ex,0.87ex) -- (1ex,0) -- (0,0);}
\def \gpru#1{\tikz\draw[scale=\pscale, color=mgray, fill={#1}, line width=0.1ex, rounded corners=0.05ex, line cap=round] (1ex,0) -- (0.75ex,0.43ex) -- (0.75ex,0.87ex) -- (0.5ex,0.87ex) -- (0,0) -- (1ex,0);}
\def \gprd#1{\tikz\draw[scale=\pscale,color=mgray, fill={#1}, line width=0.1ex, rounded corners=0.05ex, line cap=round] (1ex,0) -- (1ex,0.43ex) -- (0.75ex,0.43ex) -- (0.5ex,0.87ex) -- (0,0) -- (1ex,0);}
\def \gald#1{\tikz\draw[scale=\pscale, color=mgray, fill={#1}, line width=0.1ex, rounded corners=0.05ex, line cap=round] (0,0.58ex) -- (0.25ex,0.14ex) -- (0.25ex,-0.29ex) -- (0.5ex,-0.29ex) -- (1ex,0.58ex) -- (0,0.58ex);}
\def \galu#1{\tikz\draw[scale=\pscale, color=mgray, fill={#1}, line width=0.1ex, rounded corners=0.05ex, line cap=round] (0,0.58ex) -- (0,0.14ex) -- (0.25ex,0.14ex) -- (0.5ex,-0.29ex) -- (1ex,0.58ex) -- (0,0.58ex);}
\def \gard#1{\tikz\draw[scale=\pscale, color=mgray, fill={#1}, line width=0.1ex, rounded corners=0.05ex, line cap=round] (1ex,0.58ex) -- (0.75ex,0.14ex) -- (0.75ex,-0.29ex) -- (0.5ex,-0.29ex) -- (0,0.58ex) -- (1ex,0.58ex);}
\def \garu#1{\tikz\draw[scale=\pscale, color=mgray, fill={#1}, line width=0.1ex, rounded corners=0.05ex, line cap=round] (1ex,0.58ex) -- (1ex,0.14ex) -- (0.75ex,0.14ex) -- (0.5ex,-0.29ex) -- (0,0.58ex) -- (1ex,0.58ex);}

\def \gsqu#1{\tikz\draw[scale=\pscale, color=mgray, fill={#1}, line width=0.1ex, rounded corners=0.05ex, line cap=round] (0,0) -- (.9ex,0) -- (.9ex,.9ex) -- (0,.9ex) -- (0,0);}

\def \gcir#1{\tikz\draw[scale=\pscale, color=mgray, fill={#1}, line width=0.1ex, rounded corners=0.05ex, line cap=round] (0,0) circle (0.5ex);}

\def \ga{\gamma}

\def \ep{\epsilon}

\def \ze{\zeta}
\def \et{\eta}
\def \th{\theta}

\def \rh{\rho}
\def \si{\sigma}

\def \ch{\chi}
\def \ps{\psi}
\def \om{\omega}

\def \Ps{\Psi}

\def \beq{\begin{equation}}
\def \eeq{\end{equation}}
\def \ba{\begin{array}}
\def \ea{\end{array}}
\def \da{\dagger}
\def \pa{\partial}

\def \lb{\left[}
\def \rb{\right]}
\def \lp{\left(}
\def \rp{\right)}

\def \q{\;\;\;\;}
\def \s{\;\;\;\;\;\;}

\def \fr#1#2{{\textstyle \frac{#1}{#2}}}
\def \ha{\fr{1}{2}}
\def \nfr#1#2{\nicefrac{#1}{#2}}
\def \nha{\nfr{1}{2}}
\def \arXiv#1{\texttt{\href{http://arxiv.org/abs/#1}{arXiv:#1}}}

\def \ud#1{\underaccent{.}{{#1}}}

\def\DynkinNodeSize{3.5mm}
\def\DynkinArrowLength{3mm}
\tikzset{
  dnode/.style={
    circle,
    inner sep=0pt,
    minimum size=\DynkinNodeSize,
    fill=white,
    draw},
  middlearrow/.style={
    decoration={markings,
      mark=at position 0.6 with
      {\draw (0:0mm) -- +(+135:\DynkinArrowLength); \draw (0:0mm) -- +(-135:\DynkinArrowLength);}
    },
    postaction={decorate}
  },
  sedge/.style={},
  dedge/.style={
    middlearrow,
    double distance=0.5mm
  },
  tedge/.style={
    middlearrow,
    double distance=1.0mm+\pgflinewidth,
    postaction={draw}
  }
}

\subheader{$ $}

\title{\boldmath C, P, T, and Triality}

\author{A. Garrett Lisi}

\affiliation{Pacific Science Institute, Makawao, HI, USA}

\emailAdd{Gar@Li.si}

\abstract{Discrete charge, parity, and time symmetries (C, P, and T) of quantized fermion states are extended by a triality symmetry (t), producing the CPTt Group, transforming between three generations of fermions.}


\begin{document}

\maketitle

\newpage

\section{Introduction}

The Standard Model of particle physics has been wildly successful at explaining physical phenomenon for the past fifty years. Over that time the theory has been filled in with various particles and more accurate measurements of their interactions and masses, but the fundamental structure has endured. Perhaps the most perplexing mystery within the Standard Model is why fermions, described by Dirac spinors, exist in three generations---each generation having identical charges with respect to the fundamental forces, but different masses and mixings with respect to the Higgs field. This tri-fold symmetry of fermion generations strongly indicates there is a finite symmetry, triality, that acts in conjunction with the charge, parity, and time reversal symmetries of Dirac spinors in the Standard Model. In this work, we introduce this symmetry in the context of $C$, $P$, and $T$, and work out the group theoretical implications.

Dirac spinors are a foundational part of the Standard Model. They are not, however, an irreducible representation space of the spacetime spin group, $Spin(1,3)$. Rather, Dirac spinors are an irreducible representation space of the spacetime pin group, $Pin(1,3)$, a double cover of the spacetime orthogonal group, $O(1,3)$, and a subgroup of the spacetime Clifford group $Cl^*(1,3)$ (consisting of invertible Clifford algebra elements).{\cite{Pin}} The identity component of the spacetime spin group is the spacetime orthochronous spin group, $Spin^+(1,3)$ (the group of rotations obtained by exponentiating $Cl(1,3)$ bivectors), equivalent to $SL(2,\mathbb{C})$, and the double cover of $SO^+(1,3)$. The spacetime orthochronous spin group extends to the spacetime pin group by adding spatial reflections, called parity conjugations ($P$), and temporal reflections, called time conjugations ($T$):
$$
Spin^+(1,3)  \rtimes \{1,P,T,PT \} = Spin(1,3)  \rtimes \{1,P \} = Pin(1,3) \subset Cl^*(1,3) \subset Cl(1,3)
$$
Because the weak interaction maximally violates P-symmetry (interacting with only left-chiral parts of fermions), many authors choose to disregard foundational P-symmetry, building theories up from Weyl spinors in an irreducible representation space of $SL(2,\mathbb{C}) = Spin^+(1,3)$. However, it is very likely that all fermions consist of both left and right-chiral parts that are related by a P-symmetry that is broken but should still be considered foundational.

$Pin(1,3)$ group elements, $U$, act as $Cl(1,3)$ Clifford algebra elements on Dirac spinors, in conjunction with corresponding active Lorentz transformations, $\Ps(x) \to U \Ps(x')$, and via the Clifford adjoint on Clifford vectors and other Clifford algebra elements, $A(x) \to U A(x') \, U^-$. The $Spin^+(1,3)$ subgroup elements of $Pin(1,3)$ consist of combinations of spatial rotations and Lorentz boosts,
$$
U_\th = e^{-\ha \ga \ga_0 n_1 \th} = \cos{\fr{\th}{2}} - \ga \ga_0 n_1 \sin{\fr{\th}{2}}
\s \;\;\;\;\;
U_\ze = e^{-\ha \ga_0 n_2 \ze} = \cosh{\fr{\ze}{2}} - \ga_0 n_2 \sinh{\fr{\ze}{2}}
$$
which combine to make scalar plus bivector plus pseudoscalar ($\ga = \ga_0 \ga_1 \ga_2 \ga_3$) elements, $U \in \mbox{Spin}{}^+(1,3)$. Alternatively, Clifford rotation elements can be constructed from successive Clifford reflections. Using the pseudoscalar, the reflection of any Clifford element, $A \in Cl(1,3)$, or spinor, $\Ps$, through a vector, $u$, can also be written as a Clifford adjoint,
$$
A' = R_u A = (u \ga) A (u \ga)^- = (u \ga) A (u^- \ga)
\s \s
\Ps' = R_u \Ps = (u \ga) \Ps
$$
Arbitrary Clifford reflections, as elements of $Pin(1,3)$, can be constructed by combining spacetime rotations with a reflection through the unit time Clifford basis vector, $U'_T = \ga_0 \ga = \ga_1 \ga_2 \ga_3$, called ``unitary time conjugation'' ($T_U$), or reflection through the three unit space Clifford basis vectors, $U'_P = - \ga_1 \ga \ga_2 \ga \ga_3 \ga = \ga_0$, called ``parity conjugation'' ($P$)---with the negative sign by convention---or their combination, $U'_{PT} = \ga$. These $P$ and $T_U$ symmetry conjugations anticommute, $P T_U = - T_U P$, and close to form the $PT_U$ Group (of order 8), a finite subgroup of $Pin(1,3)$. For $Pin(1,3)$ these satisfy $\{ {U'}_P^2 = 1, {U'}_T^2 = 1, {U'}_{PT}^2 = -1 \}$, while for $Pin(3,1)$ we have $\{ U'{}_P^2 = -1, U'{}_T^2 = -1, U'{}_{PT}^2 = -1 \}$, so these two pin groups are not isomorphic. Also note that $Spin(1,3) = Spin^+(1,3) \times \{ 1, PT_U \}$. Since we like working with $Spin^+(1,3)$ because of the isomorphism to $SL(2,\mathbb{C})$, but we would also like our spinors to change sign under $P^2$, corresponding to a $2 \pi$ rotation, as they do in $Pin(3,1)$, we can fudge a bit by defining $U_P = i \ga_0$, providing the best of both worlds. Physically, we appear to live in a world with underlying $Pin(3,1)$ symmetry, but it's easier to work with $Pin(1,3)$ for calculations.

The spacetime pin group, $Pin(1,3)$, is in $Cl(1,3)$, and can be extended to the Dirac algebra via an antiunitary charge conjugation ($C$) operator, $U_C = i \ga_2 K$ (in which $K$ is the complex conjugation operator), acting on complex Dirac spinors---transforming between particles and antiparticles. When we look at quantized fermion fields, T-symmetry becomes more complicated due to the positive energy constraint. A different unitary time conjugation operator, $U^Q_T = \ga_{13}$, is associated with T-symmetry ($T$). And this is carried back to the unquantized arena by defining yet a different, antiunitary time conjugation operator on Dirac spinors, $U_T = -i U'_T U_C = \ga_{13} K$, which we also associate with T-symmetry ($T$). Under their complex conjugations and Clifford multiplications, $\{ U_C, U_P, U_T \}$ generate the $CPT$ Group (of order 16).{\cite{CPT}} The projective action of the $CPT$ Group on quantized or unquantized Dirac spinors can be graphically depicted, using weights, as action on a cube. Since the $CPT$ Group is the split-biquaternion group, it is natural to re-identify Dirac spinors as biquaternions, with corresponding $C$, $P$, and $T$ actions on them.

In physics, $P$-symmetry is violated by the weak interaction, and $CP$-symmetry is violated by the Yukawa interaction with the Higgs field, but $CPT$-symmetry holds, with $CPT \sim P T_U$ here, so $Spin(1,3) = Spin^+(1,3) \times \{1, P T_U \}$ is an unbroken symmetry of nature. In the Standard Model, Dirac spinors are found in triplets, corresponding to three generations of each kind of fermion, each combining with the others according to a mixing matrix to produce mass states. Given this three-fold symmetry, it is natural to introduce a discrete triality symmetry ($t$) that cycles between generations of Dirac spinors. The main result of this paper is that there is a unique way to nontrivially extend the $CPT$ Group by triality to the $CPTt$ Group (of order $96$) while preserving $CPT$-symmetry. This $CPTt$ Group acts projectively on a 24-cell of Dirac spinor triplets.

\newpage

\section{Gravitational Weights of a Dirac Spinor}

Fermions are quantized excitations of Dirac pinors, corresponding to the complex spinor representation space of the spacetime pin group, $Pin(1,3)$, described via the $Cl(1,3)$ Clifford algebra. Using Pauli matrices, the Weyl representation of Dirac matrices (the $Cl(1,3)$ Clifford basis vectors) are:
$$
\ga_0 = \si_1 \otimes \si_0
\s \s
\ga_\pi = - i \si_2 \otimes \si_\pi
$$
These representative matrices multiply to give six Clifford bivector basis generators of $spin(1,3)$,
$$
J_\pi = \fr{1}{4} \ep_{\pi \rh \si} \ga_{\rh \si}
= - \fr{i}{2} \si_0 \otimes \si_\pi
\s \s
K_\pi = \fr{1}{2} \ga_{0 \pi}
= \fr{1}{2} \si_3 \otimes \si_\pi
$$
corresponding to spatial rotations and spacetime boosts. Identifying the antisymmetric Clifford or matrix product of these with the Lie bracket, the nonzero brackets of the Lorentz algebra are:
$$
\left[ J_\pi, J_\rh \right] = \ep_{\pi \rh \si} J_\si \s
\left[ J_\pi, K_\rh \right] = \ep_{\pi \rh \si} K_\si \s
\left[ K_\pi, K_\rh \right] = - \ep_{\pi \rh \si} J_\si
$$
This Lie algebra structure of $spin(1,3)$ is further elucidated by identifying basis elements of a Cartan subalgebra, $\{ J_3, K_3 \}$, and computing the eigenvalues, $j_3$ and $k_3$, with imaginary and real root coordinates, $j_3^\mathbb{I}$ and $k_3^\mathbb{R}$, and corresponding root vectors, resulting in a Cartan-Weyl basis for the Lie algebra, $\{ J_3, K_3, E_L^{\vee/\wedge}, E_R^{\vee/\wedge} \}$, with nonzero brackets,
$$
\ba{rclcrclcrcl}
\lb J_3, E_L^{\vee/\wedge} \rb  & = &  (\pm i) E_L^{\vee/\wedge} & \s & \lb K_3, E_L^{\vee/\wedge} \rb  & = &  (\mp 1) E_L^{\vee/\wedge} & \s &      E_L^{\vee/\wedge}  & = &  \ha \lp \mp J_1 + i J_2 \pm i K_1 + K_2 \rp \\[5.0pt]
\lb J_3, E_R^{\vee/\wedge} \rb  & = &  (\pm i) E_R^{\vee/\wedge} & \;\; & \lb K_3, E_R^{\vee/\wedge} \rb  & = &  (\pm 1) E_R^{\vee/\wedge} & \;\; & E_R^{\vee/\wedge}  & = &  \ha \lp \pm J_1 - i J_2 \pm i K_1 + K_2 \rp \\[5.0pt]
\lb E_L^{\vee}, E_L^{\wedge} \rb  & = &   -i \, J_3 - K_3 & \;\; & \lb E_R^{\vee}, E_R^{\wedge} \rb  & = &  -i \, J_3 + K_3 & & & & \\ 
\ea
$$
These are put in Chevalley-Serre form by defining non-orthogonal Cartan basis elements, $H_{L/R} = -i J_3 \mp K_3$, resulting in the brackets,
$$
\lb H_L, E_L^{\vee/\wedge} \rb = \pm 2 i E_L^{\vee/\wedge} \s
\lb E_L^{\vee}, E_L^{\wedge} \rb = H_L \s 
\lb H_R, E_R^{\vee/\wedge} \rb = \pm 2 i E_R^{\vee/\wedge} \s
\lb E_R^{\vee}, E_R^{\wedge} \rb = H_R 
$$
Alternatively, these complex root vectors can be transformed to a real Cartan-Weyl basis, with resulting real structure constants,
$$
\ba{rclcrclcrcl}
\lb J_3,E^\mathbb{R}_{\pm} \rb  & = &  (+1) E^\mathbb{I}_{\pm} & \;\; & \lb K_3,E^\mathbb{R}_{\pm} \rb  & = &  (\pm 1) E^\mathbb{R}_{\pm} & \;\; &      E^\mathbb{R}_{\pm}  & = &   \ha \lp \pm J_1 + K_2 \rp \\[2pt] 
\lb J_3,E^\mathbb{I}_{\pm} \rb  & = &  (-1) E^\mathbb{R}_{\pm} & \;\; & \lb K_3,E^\mathbb{I}_{\pm} \rb  & = &  (\pm 1) E^\mathbb{I}_{\pm} & \;\; &      E^\mathbb{I}_{\pm}  & = &   \ha \lp \pm J_2 - K_1 \rp \\[2pt] 
\lb E^\mathbb{R}_{+},E^\mathbb{I}_{-} \rb  & = &  \lb E^\mathbb{R}_{-},E^\mathbb{I}_{+} \rb = - \ha J_3 & \s & \lb E^\mathbb{R}_{+},E^\mathbb{R}_{-} \rb  & = &  \lb E^\mathbb{I}_{+},E^\mathbb{I}_{-} \rb = \ha K_3 & \s & & & \\ 
\ea
$$

A spacetime Lorentz algebra element represented as a Clifford bivector, $B = B_s^\pi J_\pi + B_t^\pi K_\pi = \ha B^{\mu \nu} \ga_{\mu \nu}$, acts on a spacetime vector, $v = v^\mu \ga_\mu$, via antisymmetric Clifford multiplication,
$$
v' = B \times v = \fr{1}{2} B^{\mu \nu} v^\rh \lp \ga_{\mu \nu} \times \ga_\rh \rp
= \fr{1}{4} B^{\mu \nu} v^\rh \lp \ga_\mu \et_{\nu \rh} - \ga_\nu \et_{\mu \rh} \rp 
= \ha B^{\mu \nu} v_\nu \ga_\mu 
$$
Using Cartan subalgebra basis elements, $J_3 = \ha \ga_{12}$ and $K_3 = \ha \ga_{03}$, the weights and weight vectors of this spacetime vector representation space are:
$$
\ba{rclcrclcrcl}
J_3 \times v_S^{\vee/\wedge} &=& (\pm i) v_S^{\vee/\wedge} &\s& K_3 \times v_S^{\vee/\wedge} &=& 0 &\s& v_S^{\vee/\wedge} &=& \ga_1 \mp i \ga_2 \\[2pt] 
J_3 \times v_T^{\vee/\wedge} &=& 0 &\s& K_3 \times v_T^{\vee/\wedge} &=& (\pm 1) v_T^{\vee/\wedge} &\s& v_T^{\vee/\wedge} &=& \ga_0 \mp \ga_3 \\
\ea
$$
A Dirac spinor, $\Ps = \Ps^a Q_a$, is acted on by Clifford bivectors via their representative matrices, $\Ps' = B \, \Ps = \ha B^{\mu \nu} (\ga_{\mu \nu})^b{}_{c}  \Ps^c Q_b$. Using the action of the Cartan subalgebra elements, $J_3 = - \fr{i}{2} \si_0 \otimes \si_3$ and $K_3 = \fr{1}{2} \si_3 \otimes \si_3$, the weights and weight vectors of this spinor representation space of the spacetime Lorentz algebra are:
\beq
\ba{rcl c rcl c rcl c rcl}
J_3 \, \Ps_L^{\vee/\wedge} &=& (\pm \nfr{i}{2}) \Ps_L^{\vee/\wedge} &\s& K_3 \, \Ps_L^{\vee/\wedge} &=& (\mp \nha) \Ps_L^{\vee/\wedge} &\s& \Ps_L^{\wedge} &=& Q_1 &\s& \Ps_R^{\wedge} &=& Q_3  \\[2pt]
J_3 \, \Ps_R^{\vee/\wedge} &=& (\pm \nfr{i}{2}) \Ps_R^{\vee/\wedge} &\s& K_3 \, \Ps_R^{\vee/\wedge} &=& (\pm \nha) \Ps_R^{\vee/\wedge} &\s& \Ps_L^{\vee} &=& Q_2 &\s& \Ps_R^{\vee} &=& Q_4  \\
\ea
\label{spinweight}
\eeq
The roots of $spin(1,3)$ correspond to gravitational spin connection states, $\om^{\wedge/\vee}_{L/R}$, vector weights correspond to gravitational frame states, $e^{\wedge/\vee}_{S/T}$, and spinor weights correspond to massless fermion states, $f^{\wedge/\vee}_{L/R}$. Since the spin operator is $S_z = i J_3$, the corresponding spin quantum number is $\om_S = s_z =  - j_3^\mathbb{I}$; and similarly for the boost quantum number we define $\om_T = - k_3^\mathbb{R}$. (Real weight components are sometimes labeled with $\mathbb{R}$ to distinguish them from more typical imaginary weight components.)

\begin{table}[h!t]
{\centerline{\parbox{.45\textwidth}{\centerline{

\renewcommand{\arraystretch}{1.2}
\begin{tabular}
{@{\vrule width1.0pt}c@{\vrule width0.2pt}c@{\vrule width1.0pt}c@{\vrule width0.4pt}c@{\vrule width1.0pt}c@{\vrule width0.4pt}c@{\vrule width1.0pt}}
\noalign{\hrule height 1.0pt}
\multicolumn{2}{@{\vrule width1.0pt}c@{\vrule width1.0pt}}{$\;\; spin(1,3) \;\;$} & $\, k_3^\mathbb{R} \,$ & $\, j_3^\mathbb{I} \,$ & $\, \om_T \,$ & $\, \om_S \,$ \\
\noalign{\hrule height 1.0pt}
$\,$ \gcir{lgray} $\,$ & $\;\; \om^{\wedge/\vee}_L \;\;$ & \multicolumn{1}{c@{\vrule width0.0pt}}{$\!\pm1\;$} & \multicolumn{1}{c@{\vrule width1.0pt}}{$\!\mp1\;$} & \multicolumn{1}{c@{\vrule width0.0pt}}{$\!\mp1\;$} & \multicolumn{1}{c@{\vrule width1.0pt}}{$\!\pm1\;$} \\
\noalign{\hrule height 0.2pt}
$\,$ \gcir{lgray} $\,$ & $\;\; \om^{\wedge/\vee}_R \;\;$ & \multicolumn{1}{c@{\vrule width0.0pt}}{$\!\mp1\;$} & \multicolumn{1}{c@{\vrule width1.0pt}}{$\!\mp1\;$} & \multicolumn{1}{c@{\vrule width0.0pt}}{$\!\pm1\;$} & \multicolumn{1}{c@{\vrule width1.0pt}}{$\!\pm1\;$} \\
\noalign{\hrule height 1.0pt}
$\,$ \gsqu{lgray} $\,$ & $\;\; e^{\wedge/\vee}_S \;\;$ & \multicolumn{1}{c@{\vrule width0.0pt}}{$\!0\;$} & \multicolumn{1}{c@{\vrule width1.0pt}}{$\!\mp1\;$} & \multicolumn{1}{c@{\vrule width0.0pt}}{$\!0\;$} & \multicolumn{1}{c@{\vrule width1.0pt}}{$\!\pm1\;$} \\
\noalign{\hrule height 0.2pt}
$\,$ \gsqu{lgray} $\,$ & $\;\; e^{\wedge/\vee}_T \;\;$ & \multicolumn{1}{c@{\vrule width0.0pt}}{$\!\mp1\;$} & \multicolumn{1}{c@{\vrule width1.0pt}}{$\!0\;$} & \multicolumn{1}{c@{\vrule width0.0pt}}{$\!\pm1\;$} & \multicolumn{1}{c@{\vrule width1.0pt}}{$\!0\;$} \\
\noalign{\hrule height 1.0pt}
$\,$ \gplu{myell} \gpld{myell} $\,$ & $\;\; f^{\wedge/\vee}_L \;\;$ & \multicolumn{1}{c@{\vrule width0.0pt}}{$\!\pm\nha\;$} & \multicolumn{1}{c@{\vrule width1.0pt}}{$\!\mp\nha\;$} & \multicolumn{1}{c@{\vrule width0.0pt}}{$\!\mp\nha\;$} & \multicolumn{1}{c@{\vrule width1.0pt}}{$\!\pm\nha\;$} \\
\noalign{\hrule height 0.2pt}
$\,$ \gpru{myell} \gprd{myell} $\,$ & $\;\; f^{\wedge/\vee}_R \;\;$ & \multicolumn{1}{c@{\vrule width0.0pt}}{$\!\mp\nha\;$} & \multicolumn{1}{c@{\vrule width1.0pt}}{$\!\mp\nha\;$} & \multicolumn{1}{c@{\vrule width0.0pt}}{$\!\pm\nha\;$} & \multicolumn{1}{c@{\vrule width1.0pt}}{$\!\pm\nha\;$} \\
\noalign{\hrule height 1.0pt}
\end{tabular}

		}}~
		\parbox{.45\textwidth}{\centerline{
		
		\includegraphics[height=2in,width=2in]{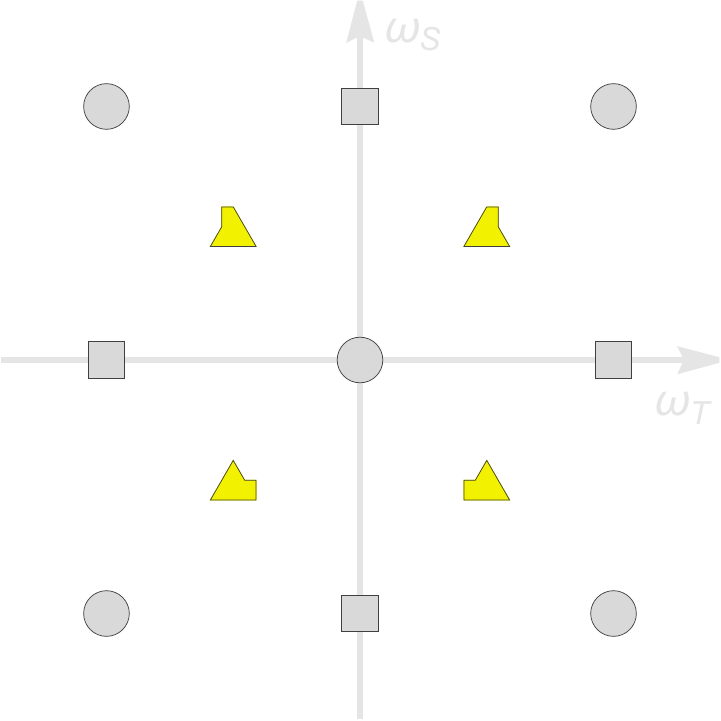}
		
			}}}
\caption{Roots and weights of $spin(1,3)$.\label{Grav}}}
\end{table}

\newpage

\section{Dirac Solutions and Identities}

A Dirac spinor solution to the Dirac equation, $0 = i \ga^\mu \pa_\mu \Ps - m \Ps$, corresponds to a fermion or antifermion with specific momentum and spin. A massless fermion can be treated in the limit, $m \to 0$, and has aligned (right-handed helicity) or anti-aligned (left-handed helicity) momentum and spin. Massless fermions can interact with a Higgs field and attain mass. For solving the Dirac equation, it is useful to express the momentum, 4-momentum, energy, and momentum direction as:
$$
p = p^\pi \si_\pi =
\lb \ba{cc}
p^3 \;&\; p^1 - i p^2 \\
p^1 + i p^2 \;&\; -p^3
\ea \rb
\s \s \s
\ba{c}
p^\mu = \lp E, p^1, p^2, p^3 \rp \\[3pt]
E^2 = p^2 + m^2 \\
p = \left| p \right| p_u
\ea
$$
Dirac solutions consist of positive and negative energy parts, $\Ps_+ = u_p^{\wedge/\vee} e^{- i p_\mu x^\mu}$ and  $\Ps_- = v_p^{\wedge/\vee} e^{+ i p_\mu x^\mu}$, which can be expressed as rest spinors Lorentz boosted to momentum, $p$,
$$
\ba{rcl}

u_p^{\wedge/\vee} &=& U_p u_0^{\wedge/\vee}
= \lb \ba{cc} \cosh{\fr{\ze}{2}} - p_u \sinh{\fr{\ze}{2}} & 0 \\ 0 & \cosh{\fr{\ze}{2}} + p_u \sinh{\fr{\ze}{2}} \ea \rb
 \fr{1}{\sqrt{2}} \lb \ba{c} \ch^{\wedge/\vee} \\ \ch^{\wedge/\vee} \ea \rb
 =  \fr{1}{\sqrt{2m}} \lb \ba{c} \sqrt{E-p} \; \ch^{\wedge/\vee} \\ \sqrt{E+p} \; \ch^{\wedge/\vee} \ea \rb \\[20pt]

v_p^{\wedge/\vee} &=& U_p v_0^{\wedge/\vee}
= \lb \ba{cc} \cosh{\fr{\ze}{2}} - p_u \sinh{\fr{\ze}{2}} & 0 \\ 0 & \cosh{\fr{\ze}{2}} + p_u \sinh{\fr{\ze}{2}} \ea \rb
 \fr{1}{\sqrt{2}} \lb \ba{c} \xi^{\wedge/\vee} \\ -\xi^{\wedge/\vee} \ea \rb
 =  \fr{1}{\sqrt{2m}} \lb \ba{c} \sqrt{E-p} \; \xi^{\wedge/\vee} \\ -\sqrt{E+p} \; \xi^{\wedge/\vee} \ea \rb \\

\ea
$$
in which the rapidity, $\ze$, satisfies, $\{ \sinh{\ze} = \fr{\left| p \right|}{m}, \cosh{\ze} = \fr{E}{m}, \tanh{\ze} = \fr{\left| p \right|}{E} \}$, $\ch^{\wedge/\vee}$ are up or down unit Pauli spinors, and $\xi^{\wedge/\vee} = \ep \ch^{\wedge/\vee} =  \pm \ch^{\vee/\wedge}$ are the flipped unit spinors, in which $\ep = -i \si_2$ is the skew matrix. These solutions are normalized, $\bar{u} u = 1 = - \bar{v} v$. In the massless limit, $m \to 0$, we have $\cosh{\fr{\ze}{2}} = \sinh{\fr{\ze}{2}}$ and therefore:
\beq
u_{p0}^{\wedge/\vee} =  \ha \lb \ba{c} \lp 1-p_u \rp \ch^{\wedge/\vee} \\ \lp 1+p_u \rp \ch^{\wedge/\vee} \ea \rb \s \s
v_{p0}^{\wedge/\vee} =  \ha \lb \ba{c} \lp 1-p_u \rp \xi^{\wedge/\vee} \\ - \lp 1+p_u \rp \xi^{\wedge/\vee} \ea \rb
\label{massless}
\eeq
which satisfy $\bar{u} u = \bar{v} v = 0$ and have been re-normalized so that $u^{\dagger} u = v^{\dagger} v = 1$.

Using our Dirac matrices,
$$
i \ga_2 = \lb \ba{cc} 0 \;&\; \ep \\ -\ep \;&\; 0 \ea \rb
\s \s \s
i \ga_0 = \lb \ba{cc} 0 \;&\; i \\ i \;&\; 0 \ea \rb
\s \s \s
\ga_{13} = \lb \ba{cc}  -\ep & 0 \\ 0 &-\ep \ea \rb
\s
$$
and a Pauli matrix identity, $\si_\pi^* = \ep \, \si_\pi \ep$, we have a collection of Dirac solution identities relevant to $C$, $P$, and $T$ symmetries:
\beq
\ba{rclcrclcrcl}

i \ga_2  \, u_p^{\wedge/\vee \, *}  &=&  v_p^{\wedge/\vee} & \s \s &  i \ga_0 \, u_{-p}^{\wedge/\vee} &=&  + i \, u_p^{\wedge/\vee}  & \s \s & \ga_{13}  \, u_{-p}^{\wedge/\vee}  &=& \mp \,  u_p^{\vee/\wedge \, *} \\[10pt]

i \ga_2  \, v_p^{\wedge/\vee \, *}  &=&  u_p^{\wedge/\vee} & \s \s & i \ga_0 \, v_{-p}^{\wedge/\vee} &=& - i \, v_p^{\wedge/\vee} & \s \s & \ga_{13} \, v_{-p}^{\wedge/\vee}  &=& \mp  \,  v_p^{\vee/\wedge \, *}

\ea
\label{did}
\eeq
Note that these identities also hold for the massless case.

Our massless Dirac solutions, (\ref{massless}), are not eigenstates of helicity, $h = \ha \si_0 \otimes p_u$, unless the momentum is along or counter to the $z$ axis, $p_u = \pm \si_3$, in which cases we can relate the solutions to spin weight vectors, (\ref{spinweight}), with $L/R$ denoting left or right helicity, $\mp \ha$,
$$
\ba{cccc}
u_{-0}^{\wedge} =  \lb \ba{c} 1 \\ 0 \\ 0 \\ 0 \ea \rb = \Ps_L^\wedge \s &
u_{+0}^{\vee} =  \lb \ba{c} 0 \\ 1 \\ 0 \\ 0 \ea \rb = \Ps_L^\vee \s &
u_{+0}^{\wedge} =  \lb \ba{c} 0 \\ 0 \\ 1 \\ 0 \ea \rb = \Ps_R^\wedge \s &
u_{-0}^{\vee} =  \lb \ba{c} 0 \\ 0 \\ 0 \\ 1 \ea \rb = \Ps_R^\vee \\[45pt]

v_{-0}^{\wedge} =  \lb \ba{c} 0 \\ 0 \\ 0 \\ -1 \ea \rb = -\Ps_R^\vee \s &
v_{+0}^{\vee} =  \lb \ba{c} 0 \\ 0 \\ 1 \\ 0 \ea \rb = \Ps_R^\wedge \s &
v_{+0}^{\wedge} =  \lb \ba{c} 0 \\ 1 \\ 0 \\ 0 \ea \rb = \Ps_L^\vee \s &
v_{-0}^{\vee} =  \lb \ba{c} -1 \\ 0 \\ 0 \\ 0 \ea \rb = -\Ps_L^\wedge
\ea
$$
Note that the massless negative energy solutions, $v^{\wedge/\vee}_{\pm 0}$, have exactly the opposite spin and helicity one might expect from their symbology.  

Applied to massless fermions with momentum along or counter to $\hat{z}$, charge ($C$) conjugation transforms to antifermions with the same spin, helicity, and momentum, parity ($P$) conjugation transforms to fermions with the same spin, opposite helicity and momentum, and time ($T$) conjugation transforms to fermions with opposite spin, same helicity, and opposite momentum. To see the precise phases involved in these conjugations, we need to consider their actions on massless quantized Dirac fermions.

$$
\s
$$

\begin{center}
\begin{tikzpicture}[scale=1.0]
  \coordinate (A) at (-1.5,.8);
  \coordinate (B) at (1.5,.8);
  \coordinate (C) at (-1.5,-.8);
  \coordinate (D) at (1.5,-.8);
  
  \coordinate (E) at (.05,.7);
  \coordinate (F) at (.05,-.12);
  \coordinate (G) at (.05,-.92);

\draw[<->, thick, >=latex, line width=0.6pt] (E) ++(-11pt,0) -- ++(+20pt,0);
\node[above=0pt] at (E) {$P$};

\draw[<->, thick, >=latex, line width=0.6pt] (G) ++(-11pt,0) -- ++(+20pt,0);
\node[above=0pt] at (G) {$P$};

\draw[<->, thick, >=latex, line width=0.6pt] (F) ++(-11pt,7pt) -- ++(+20pt,-14pt);
\draw[<->, thick, >=latex, line width=0.6pt] (F) ++(-11pt,-7pt) -- ++(+20pt,+14pt);
\node[above=3pt] at (F) {$T$};
  
  \node[scale=2.0] at (A) {\gplu{myell}};
  \draw[->, thick, >=latex, line width=1.0pt] (A) ++(0,-12pt) -- ++(0,-14pt);
  \draw[thick, <-, >=latex, line width=.8pt] (A)+(10pt,2pt) arc[start angle=60, end angle=-240, x radius=20pt, y radius=6pt];
  \node[left=25pt] at (A) {$\Ps_L^\wedge$};
 
  \node[scale=2.0] at (B) {\gpru{myell}};
  \draw[->, thick, >=latex, line width=1.0pt] (B) ++(0,10pt) -- ++(0,14pt);
  \draw[thick, <-, >=latex, line width=.8pt] (B)+(10pt,2pt) arc[start angle=60, end angle=-240, x radius=20pt, y radius=6pt];
  \node[right=25pt] at (B) {$\Ps_R^\wedge$};
 
  \node[scale=2.0] at (C) {\gprd{myell}};
  \draw[->, thick, >=latex, line width=1.0pt] (C) ++(0,-12pt) -- ++(0,-14pt);
  \draw[thick, ->, >=latex, line width=.8pt] (C)+(11pt,2pt) arc[start angle=60, end angle=-240, x radius=20pt, y radius=6pt];
  \node[left=25pt] at (C) {$\Ps_R^\vee$};
 
  \node[scale=2.0] at (D) {\gpld{myell}};
  \draw[->, thick, >=latex, line width=1.0pt] (D) ++(0,10pt) -- ++(0,14pt);
  \draw[thick, ->, >=latex, line width=.8pt] (D)+(10pt,2pt) arc[start angle=60, end angle=-240, x radius=20pt, y radius=6pt];
  \node[right=25pt] at (D) {$\Ps_L^\vee$};
\end{tikzpicture}
\end{center}

\newpage

\section{Gravitational Weights of a Massless Quantum Dirac Spinor}

A quantum Dirac spinor in Minkowski spacetime, 
$$
\ud{\hat{\Ps}} = \int{\fr{d^3p}{(2\pi)^3 (2E)}} \lp \ud{\hat{a}}_{\, p}^{\wedge/\vee} u_p^{\wedge/\vee} e^{-i p_\mu x^\mu} + \ud{\hat{b}}_{\, p}^{\wedge/\vee \, \da} v_p^{\wedge/\vee} e^{+i p_\mu x^\mu} \rp
$$
and its adjoint,
$$
\ud{\hat{\bar{\Psi}}} = \ud{\hat{\Psi}}^\da \ga^0
= \int{\fr{d^3p}{(2\pi)^3 (2E)}} \lp \ud{\hat{a}}_p^{\wedge/\vee \, \da} {\bar{u}}_p^{\wedge/\vee} e^{+i p_\mu x^\mu} + \ud{\hat{b}}_p^{\wedge/\vee} {\bar{v}}_p^{\wedge/\vee} e^{-i p_\mu x^\mu} \rp
$$
include creation and annihilation operators for particles, $\ud{\hat{a}}_{\, p}^{\wedge/\vee}$, and antiparticles, $\ud{\hat{b}}_{\, p}^{\wedge/\vee}$, of up and down spin, for all possible momenta. If we consider only the basis states of momenta in the positive, $+\hat{z}$, or negative, $-\hat{z}$, direction, the massless quantum Dirac spinor along $z$ is:
\beq
\ud{\hat{\Ps}}_z =
\lb
\ba{c}
 \ud{\hat{a}}_{-}^{\wedge} e^{-i E (t + z)} - \ud{\hat{b}}_{-}^{\vee \, \da} e^{+i E (t + z)} \\
 \ud{\hat{a}}_{+}^{\vee} e^{-i E (t - z)} + \ud{\hat{b}}_{+}^{\wedge \, \da} e^{+i E (t - z)} \\
 \ud{\hat{a}}_{+}^{\wedge} e^{-i E (t - z)} + \ud{\hat{b}}_{+}^{\vee \, \da} e^{+i E (t - z)} \\
 \ud{\hat{a}}_{-}^{\vee} e^{-i E (t + z)} - \ud{\hat{b}}_{-}^{\wedge \, \da} e^{+i E (t + z)} \\
\ea
\rb
\label{QD}
\eeq
Similarly, the massless quantum Dirac spinor adjoint along $z$ is:
\beq
\ud{\hat{\bar{\Psi}}}_z =
\lb
\ba{cccc}
\ud{\hat{a}}_{+}^{\wedge \, \da} e^{+-} + \ud{\hat{b}}_{+}^{\vee} e^{--} \q &
\ud{\hat{a}}_{-}^{\vee \, \da} e^{++} - \ud{\hat{b}}_{-}^{\wedge} e^{-+} \q &
\ud{\hat{a}}_{-}^{\wedge \, \da} e^{++} - \ud{\hat{b}}_{-}^{\vee} e^{-+} \q &
\ud{\hat{a}}_{+}^{\vee \, \da} e^{+-} + \ud{\hat{b}}_{+}^{\wedge} e^{--} 
\ea
\rb
\label{QDC}
\eeq
These are acted upon by the Lorentz algebra, $spin(1,3)$, with Cartan subalgebra basis elements chosen to be the (anti-Hermitian) spin, $J_3 = \ha \ga_{12} = -\fr{i}{2} \si_0 \otimes \si_3 = -i \, S_z$, and (Hermitian) boost, $K_3 = \ha \ga_{03} = \ha \si_3 \otimes \si_3$, bivectors of the $Cl(1,3)$ Clifford algebra. Typically, such as for the anti-Hermitian rotation operator, $O = J_3$, there is a corresponding anti-Hermitian operator on the infinite-dimensional unitary representation space operators of Quantum Field Theory, such as $\hat{O} = \hat{J}_3$, satisfying:
\beq
\lb \hat{O}, \ud{\hat{\Ps}} \rb = O \, \ud{\hat{\Ps}} \s \s
\lb \hat{J}_3, \ud{\hat{\Ps}} \rb = J_3 \, \ud{\hat{\Ps}}
\label{JP}
\eeq
and, for the adjoint,
\beq
\lb \hat{O}^\da, \ud{\hat{\bar{\Psi}}} \rb = - \ud{\hat{\bar{\Psi}}}  \lp \ga_0 O^\da \ga^0 \rp \s \s
\lb \hat{J}_3, \ud{\hat{\bar{\Psi}}} \rb = - \ud{\hat{\bar{\Psi}}} \, J_3 
\label{PJ}
\eeq
For the Hermitian boost operator, $K_3$, we take the corresponding operator, $\hat{K}_3$ on the infinite-dimensional unitary representation space to be anti-Hermitian to preserve quantum unitarity, and so we have:
\beq
\lb \hat{K}_3, \ud{\hat{\Ps}} \rb = K_3 \, \ud{\hat{\Ps}} \s \s
\lb \hat{K}_3, \ud{\hat{\bar{\Psi}}} \rb = \ud{\hat{\bar{\Psi}}} \, \lp \ga_0 K_3^\da \ga^0 \rp = - \ud{\hat{\bar{\Psi}}} \, K_3 
\label{KP}
\eeq
These formulas allow us to find the spin and boost eigenvalues (weights), $j_3$ and $k_3$, of the fermion annihilation and creation operators, using $J_3 = -\fr{i}{2} \si_0 \otimes \si_3$, $K_3 = \ha \si_3 \otimes \si_3$, and (\ref{QD}, \ref{QDC}, \ref{JP}, \ref{PJ}, \ref{KP}),
$$
\ba{rclcrclcrclcrclc}
\lb \hat{K}_3, \ud{\hat{a}}_{-}^{\wedge} \rb  &\!=\!&  \lp +\nfr{1}{2} \rp \ud{\hat{a}}_{-}^{\wedge} & \;\; &
\lb \hat{J}_3, \ud{\hat{a}}_{-}^{\wedge} \rb  &\!=\!&   \lp -\nfr{i}{2} \rp  \ud{\hat{a}}_{-}^{\wedge} & \;\; &
\lb \hat{K}_3, \ud{\hat{b}}_{-}^{\vee \, \da} \rb  &\!=\!&  \lp +\nfr{1}{2} \rp \ud{\hat{b}}_{-}^{\vee \, \da} & \;\; &
\lb \hat{J}_3, \ud{\hat{b}}_{-}^{\vee \, \da} \rb  &\!=\!& \lp -\nfr{i}{2} \rp  \ud{\hat{b}}_{-}^{\vee \, \da} \\[5pt]

\lb \hat{K}_3, \ud{\hat{a}}_{+}^{\vee} \rb  &\!=\!&  \lp -\nfr{1}{2} \rp \ud{\hat{a}}_{+}^{\vee} & \;\; &
\lb \hat{J}_3, \ud{\hat{a}}_{+}^{\vee} \rb  &\!=\!&   \lp +\nfr{i}{2} \rp  \ud{\hat{a}}_{+}^{\vee} & \;\; &
\lb \hat{K}_3, \ud{\hat{b}}_{+}^{\wedge \, \da} \rb  &\!=\!&  \lp -\nfr{1}{2} \rp \ud{\hat{b}}_{+}^{\wedge \, \da} & \;\; &
\lb \hat{J}_3, \ud{\hat{b}}_{+}^{\wedge \, \da} \rb  &\!=\!& \lp +\nfr{i}{2} \rp  \ud{\hat{b}}_{+}^{\wedge \, \da} \\[5pt]

\lb \hat{K}_3, \ud{\hat{a}}_{+}^{\wedge} \rb  &\!=\!&  \lp -\nfr{1}{2} \rp \ud{\hat{a}}_{+}^{\wedge} & \;\; &
\lb \hat{J}_3, \ud{\hat{a}}_{+}^{\wedge} \rb  &\!=\!&   \lp -\nfr{i}{2} \rp  \ud{\hat{a}}_{+}^{\wedge} & \;\; &
\lb \hat{K}_3, \ud{\hat{b}}_{+}^{\vee \, \da} \rb  &\!=\!&  \lp -\nfr{1}{2} \rp \ud{\hat{b}}_{+}^{\vee \, \da} & \;\; &
\lb \hat{J}_3, \ud{\hat{b}}_{+}^{\vee \, \da} \rb  &\!=\!& \lp -\nfr{i}{2} \rp  \ud{\hat{b}}_{+}^{\vee \, \da} \\[5pt]

\lb \hat{K}_3, \ud{\hat{a}}_{-}^{\vee} \rb  &\!=\!&  \lp +\nfr{1}{2} \rp \ud{\hat{a}}_{-}^{\vee} & \;\; &
\lb \hat{J}_3, \ud{\hat{a}}_{-}^{\vee} \rb  &\!=\!&   \lp +\nfr{i}{2} \rp  \ud{\hat{a}}_{-}^{\vee} & \;\; &
\lb \hat{K}_3, \ud{\hat{b}}_{-}^{\wedge \, \da} \rb  &\!=\!&  \lp +\nfr{1}{2} \rp \ud{\hat{b}}_{-}^{\wedge \, \da} & \;\; &
\lb \hat{J}_3, \ud{\hat{b}}_{-}^{\wedge \, \da} \rb  &\!=\!& \lp +\nfr{i}{2} \rp  \ud{\hat{b}}_{-}^{\wedge \, \da} \\[15pt]

\lb \hat{K}_3, \ud{\hat{a}}_{+}^{\wedge \, \da} \rb  &\!=\!&  \lp -\nfr{1}{2} \rp \ud{\hat{a}}_{+}^{\wedge \, \da} & \;\; &
\lb \hat{J}_3, \ud{\hat{a}}_{+}^{\wedge \, \da} \rb  &\!=\!&   \lp +\nfr{i}{2} \rp  \ud{\hat{a}}_{+}^{\wedge \, \da} & \;\; &
\lb \hat{K}_3, \ud{\hat{b}}_{+}^{\vee} \rb  &\!=\!&  \lp -\nfr{1}{2} \rp \ud{\hat{b}}_{+}^{\vee} & \;\; &
\lb \hat{J}_3, \ud{\hat{b}}_{+}^{\vee} \rb  &\!=\!& \lp +\nfr{i}{2} \rp  \ud{\hat{b}}_{+}^{\vee} \\[5pt]

\lb \hat{K}_3, \ud{\hat{a}}_{-}^{\vee \, \da} \rb  &\!=\!&  \lp +\nfr{1}{2} \rp \ud{\hat{a}}_{-}^{\vee \, \da} & \;\; &
\lb \hat{J}_3, \ud{\hat{a}}_{-}^{\vee \, \da} \rb  &\!=\!&   \lp -\nfr{i}{2} \rp  \ud{\hat{a}}_{-}^{\vee \, \da} & \;\; &
\lb \hat{K}_3, \ud{\hat{b}}_{-}^{\wedge} \rb  &\!=\!&  \lp +\nfr{1}{2} \rp \ud{\hat{b}}_{-}^{\wedge} & \;\; &
\lb \hat{J}_3, \ud{\hat{b}}_{-}^{\wedge} \rb  &\!=\!& \lp -\nfr{i}{2} \rp  \ud{\hat{b}}_{-}^{\wedge} \\[5pt]

\lb \hat{K}_3, \ud{\hat{a}}_{-}^{\wedge \, \da} \rb  &\!=\!&  \lp +\nfr{1}{2} \rp \ud{\hat{a}}_{-}^{\wedge \, \da} & \;\; &
\lb \hat{J}_3, \ud{\hat{a}}_{-}^{\wedge \, \da} \rb  &\!=\!&   \lp +\nfr{i}{2} \rp  \ud{\hat{a}}_{-}^{\wedge \, \da} & \;\; &
\lb \hat{K}_3, \ud{\hat{b}}_{-}^{\vee} \rb  &\!=\!&  \lp +\nfr{1}{2} \rp \ud{\hat{b}}_{-}^{\vee} & \;\; &
\lb \hat{J}_3, \ud{\hat{b}}_{-}^{\vee} \rb  &\!=\!& \lp +\nfr{i}{2} \rp  \ud{\hat{b}}_{-}^{\vee} \\[5pt]

\lb \hat{K}_3, \ud{\hat{a}}_{+}^{\vee \, \da} \rb  &\!=\!&  \lp -\nfr{1}{2} \rp \ud{\hat{a}}_{+}^{\vee \, \da} & \;\; &
\lb \hat{J}_3, \ud{\hat{a}}_{+}^{\vee \, \da} \rb  &\!=\!&   \lp -\nfr{i}{2} \rp  \ud{\hat{a}}_{+}^{\vee \, \da} & \;\; &
\lb \hat{K}_3, \ud{\hat{b}}_{+}^{\wedge} \rb  &\!=\!&  \lp -\nfr{1}{2} \rp \ud{\hat{b}}_{+}^{\wedge} & \;\; &
\lb \hat{J}_3, \ud{\hat{b}}_{+}^{\wedge} \rb  &\!=\!& \lp -\nfr{i}{2} \rp  \ud{\hat{b}}_{+}^{\wedge} \\
\ea
$$
Summarizing this structure, the table of spin and boost quantum numbers, $\om_S = s_z = - j_3^\mathbb{I} $ and $\om_T = - k_3^\mathbb{R}$, of the annihilation and creation operators of a massless quantum Dirac spinor along $z$, with a relabeling to particle spin and helicity, $a^{\wedge/\vee}_{\pm} \leftrightarrow a^{\wedge/\vee}_{L/R}$, is:

\begin{table}[h!t]
\centering
\renewcommand{\arraystretch}{1.1}
\begin{tabular}
{@{\vrule width1.0pt}c@{\vrule width0.2pt}c@{\vrule width0.2pt}c@{\vrule width1.0pt}c@{\vrule width0.4pt}c@
{\vrule width1.0pt}c@{\vrule width1.0pt}c@{\vrule width1.0pt}}
\noalign{\hrule height 1.0pt}
\multicolumn{3}{@{\vrule width1.0pt}c@{\vrule width1.0pt}}{$ 4_s $} & $\, \om_T \,$ & $\, \om_S \,$ & $\, h \,$ & $\; q \;$ \\

\noalign{\hrule height 1.0pt}
$\,$ \gplu{myell} $\,$ & $\;\; a^\wedge_L \;\;$ & $\;\; \hat{\ud{a}}^\wedge_- \;\;$ & \multicolumn{1}{c@{\vrule width0.0pt}}{$\!-\nha\;$} & \multicolumn{1}{c@{\vrule width1.0pt}}{$\!+\nha\;$} & \multicolumn{1}{c@{\vrule width1.0pt}}{$\!-\nha\;$} & \multicolumn{1}{c@{\vrule width1.0pt}}{$\!+\nha\;$} \\
\noalign{\hrule height 0.2pt}
$\,$ \gpld{myell} $\,$ & $\; a^\vee_L \;$ & $\; \hat{\ud{a}}^\vee_+ \;$ & \multicolumn{1}{c@{\vrule width0.0pt}}{$\!+\nha\;$} & \multicolumn{1}{c@{\vrule width1.0pt}}{$\!-\nha\;$} & \multicolumn{1}{c@{\vrule width1.0pt}}{$\!-\nha\;$} & \multicolumn{1}{c@{\vrule width1.0pt}}{$\!+\nha\;$} \\
\noalign{\hrule height 0.2pt}
$\,$ \gpru{myell} $\,$ & $\; a^\wedge_R \;$ & $\; \hat{\ud{a}}^\wedge_+ \;$ & \multicolumn{1}{c@{\vrule width0.0pt}}{$\!+\nha\;$} & \multicolumn{1}{c@{\vrule width1.0pt}}{$\!+\nha\;$} & \multicolumn{1}{c@{\vrule width1.0pt}}{$\!+\nha\;$} & \multicolumn{1}{c@{\vrule width1.0pt}}{$\!+\nha\;$} \\
\noalign{\hrule height 0.2pt}
$\,$ \gprd{myell} $\,$ & $\; a^\vee_R \;$ & $\; \hat{\ud{a}}^\vee_- \;$ & \multicolumn{1}{c@{\vrule width0.0pt}}{$\!-\nha\;$} & \multicolumn{1}{c@{\vrule width1.0pt}}{$\!-\nha\;$} & \multicolumn{1}{c@{\vrule width1.0pt}}{$\!+\nha\;$} & \multicolumn{1}{c@{\vrule width1.0pt}}{$\!+\nha\;$} \\
\noalign{\hrule height 1.0pt}

$\,$ \galu{myell} $\,$ & $\; \bar{a}^\wedge_L \;$ & $\; \hat{\ud{b}}^\wedge_- \;$ & \multicolumn{1}{c@{\vrule width0.0pt}}{$\!-\nha\;$} & \multicolumn{1}{c@{\vrule width1.0pt}}{$\!+\nha\;$} & \multicolumn{1}{c@{\vrule width1.0pt}}{$\!-\nha\;$} & \multicolumn{1}{c@{\vrule width1.0pt}}{$\!-\nha\;$} \\
\noalign{\hrule height 0.2pt}
$\,$ \gald{myell} $\,$ & $\; \bar{a}^\vee_L{} \;$ & $\; \hat{\ud{b}}^\vee_+ \;$ & \multicolumn{1}{c@{\vrule width0.0pt}}{$\!+\nha\;$} & \multicolumn{1}{c@{\vrule width1.0pt}}{$\!-\nha\;$} & \multicolumn{1}{c@{\vrule width1.0pt}}{$\!-\nha\;$} & \multicolumn{1}{c@{\vrule width1.0pt}}{$\!-\nha\;$} \\
\noalign{\hrule height 0.2pt}
$\,$ \garu{myell} $\,$ & $\; \bar{a}^\wedge_R \;$ & $\; \hat{\ud{b}}^\wedge_+ \;$ & \multicolumn{1}{c@{\vrule width0.0pt}}{$\!+\nha\;$} & \multicolumn{1}{c@{\vrule width1.0pt}}{$\!+\nha\;$} & \multicolumn{1}{c@{\vrule width1.0pt}}{$\!+\nha\;$} & \multicolumn{1}{c@{\vrule width1.0pt}}{$\!-\nha\;$} \\
\noalign{\hrule height 0.2pt}
$\,$ \gard{myell} $\,$ & $\; \bar{a}^\vee_R \;$ & $\; \hat{\ud{b}}^\vee_- \;$ & \multicolumn{1}{c@{\vrule width0.0pt}}{$\!-\nha\;$} & \multicolumn{1}{c@{\vrule width1.0pt}}{$\!-\nha\;$} & \multicolumn{1}{c@{\vrule width1.0pt}}{$\!+\nha\;$} & \multicolumn{1}{c@{\vrule width1.0pt}}{$\!-\nha\;$} \\
\noalign{\hrule height 1.0pt}
\end{tabular}
\s \s \s
\begin{tabular}
{@{\vrule width1.0pt}c@{\vrule width0.2pt}c@{\vrule width1.0pt}c@{\vrule width0.4pt}c@
{\vrule width1.0pt}c@{\vrule width1.0pt}c@{\vrule width1.0pt}}
\noalign{\hrule height 1.0pt}
\multicolumn{2}{@{\vrule width1.0pt}c@{\vrule width1.0pt}}{$\, 4_s^\da \,$} & $\, \om_T \,$ & $\, \om_S \,$ & $\, h \,$ & $\; q \;$ \\

\noalign{\hrule height 1.0pt}
$\;\; a^\wedge_L{}^\da \;\;$ & $\;\; \hat{\ud{a}}^\wedge_-{}^\da \;\;$ & \multicolumn{1}{c@{\vrule width0.0pt}}{$\!-\nha\;$} & \multicolumn{1}{c@{\vrule width1.0pt}}{$\!-\nha\;$} & \multicolumn{1}{c@{\vrule width1.0pt}}{$\!+\nha\;$} & \multicolumn{1}{c@{\vrule width1.0pt}}{$\!-\nha\;$} \\
\noalign{\hrule height 0.2pt}
$\; a^\vee_L{}^\da \;$ & $\; \hat{\ud{a}}^\vee_+{}^\da \;$ & \multicolumn{1}{c@{\vrule width0.0pt}}{$\!+\nha\;$} & \multicolumn{1}{c@{\vrule width1.0pt}}{$\!+\nha\;$} & \multicolumn{1}{c@{\vrule width1.0pt}}{$\!+\nha\;$} & \multicolumn{1}{c@{\vrule width1.0pt}}{$\!-\nha\;$} \\
\noalign{\hrule height 0.2pt}
$\; a^\wedge_R{}^\da \;$ & $\; \hat{\ud{a}}^\wedge_+{}^\da \;$ & \multicolumn{1}{c@{\vrule width0.0pt}}{$\!+\nha\;$} & \multicolumn{1}{c@{\vrule width1.0pt}}{$\!-\nha\;$} & \multicolumn{1}{c@{\vrule width1.0pt}}{$\!-\nha\;$} & \multicolumn{1}{c@{\vrule width1.0pt}}{$\!-\nha\;$} \\
\noalign{\hrule height 0.2pt}
$\; a^\vee_R{}^\da \;$ & $\; \hat{\ud{a}}^\vee_-{}^\da \;$ & \multicolumn{1}{c@{\vrule width0.0pt}}{$\!-\nha\;$} & \multicolumn{1}{c@{\vrule width1.0pt}}{$\!+\nha\;$} & \multicolumn{1}{c@{\vrule width1.0pt}}{$\!-\nha\;$} & \multicolumn{1}{c@{\vrule width1.0pt}}{$\!-\nha\;$} \\
\noalign{\hrule height 1.0pt}

$\; \bar{a}^\wedge_L{}^\da \;$ & $\; \hat{\ud{b}}^\wedge_-{}^\da \;$ & \multicolumn{1}{c@{\vrule width0.0pt}}{$\!-\nha\;$} & \multicolumn{1}{c@{\vrule width1.0pt}}{$\!-\nha\;$} & \multicolumn{1}{c@{\vrule width1.0pt}}{$\!+\nha\;$} & \multicolumn{1}{c@{\vrule width1.0pt}}{$\!+\nha\;$} \\
\noalign{\hrule height 0.2pt}
$\; \bar{a}^\vee_L{}^\da \;$ & $\; \hat{\ud{b}}^\vee_+{}^\da \;$ & \multicolumn{1}{c@{\vrule width0.0pt}}{$\!+\nha\;$} & \multicolumn{1}{c@{\vrule width1.0pt}}{$\!+\nha\;$} & \multicolumn{1}{c@{\vrule width1.0pt}}{$\!+\nha\;$} & \multicolumn{1}{c@{\vrule width1.0pt}}{$\!+\nha\;$} \\
\noalign{\hrule height 0.2pt}
$\; \bar{a}^\wedge_R{}^\da \;$ & $\; \hat{\ud{b}}^\wedge_+{}^\da \;$ & \multicolumn{1}{c@{\vrule width0.0pt}}{$\!+\nha\;$} & \multicolumn{1}{c@{\vrule width1.0pt}}{$\!-\nha\;$} & \multicolumn{1}{c@{\vrule width1.0pt}}{$\!-\nha\;$} & \multicolumn{1}{c@{\vrule width1.0pt}}{$\!+\nha\;$} \\
\noalign{\hrule height 0.2pt}
$\; \bar{a}^\vee_R{}^\da \;$ & $\; \hat{\ud{b}}^\vee_-{}^\da \;$ & \multicolumn{1}{c@{\vrule width0.0pt}}{$\!-\nha\;$} & \multicolumn{1}{c@{\vrule width1.0pt}}{$\!+\nha\;$} & \multicolumn{1}{c@{\vrule width1.0pt}}{$\!-\nha\;$} & \multicolumn{1}{c@{\vrule width1.0pt}}{$\!+\nha\;$} \\
\noalign{\hrule height 1.0pt}
\end{tabular}
\caption{The weights of the annihilation and creation operators of a charged massless quantum Dirac spinor, $4_s$, of $spin(1,3)$.}
\label{table:p4s}
\end{table}

\noindent The helicity quantum number is $h \!=\! p_z s_z = 2 \, \om_T \, \om_S$, and the $q = \pm \nha$ quantum number is for whatever internal charge the particle has. Note that the helicity, spin, and charge, but not boost, quantum numbers of a creation operator are opposite that of the corresponding annihilation operator; and there is a weight match between annihilating particles and creating antiparticles, such as $\bar{a}_L^{\wedge \, \da} = a_R^{\vee}$. The spin and boost quantum numbers of the annihilation of a massless fermion match those of a Dirac spinor (Table \ref{Grav}), as do those of the corresponding antifermion.

\newpage

\section{Charge, Parity, Time, and the CPT Group}

The charge, parity, and time conjugates of Dirac solutions are also Dirac solutions, and correspond to conjugations of a quantum Dirac spinor. A faithful representative group operator, $o(g)$, on Dirac spinors has a corresponding operator, $\hat{o}(g)$, on the infinite-dimensional representation space operators (i.e. creation and annihilation operators) of QFT,
$$
\hat{\ud{\Ps}}^g(x) = \hat{o}(g) \, \hat{\ud{\Ps}}(x) \, \hat{o}^-(g) = o(g) \, \hat{\ud{\Ps}}(x'_g(x))
$$

The antiunitary charge conjugate, $\Ps^C = i \ga_2 \Ps^*$, is:
$$
\ba{rcl}
\ud{\hat{\Ps}}^C
 &=& \int{\fr{d^3p}{(2\pi)^3 (2E)}} \lp  \lp \ud{\hat{a}}_{\, p}^{\wedge/\vee} \rp^C u_p^{\wedge/\vee} e^{-i p_\mu x^\mu} + \lp \ud{\hat{b}}_{\, p}^{\wedge/\vee \, \da} \rp^C v_p^{\wedge/\vee} e^{+i p_\mu x^\mu} \rp \\[5pt]
 &=& i \ga_2 \int{\fr{d^3p}{(2\pi)^3 (2E)}} \lp  \ud{\hat{a}}_{\, p}^{\wedge/\vee} u_p^{\wedge/\vee} e^{-i p_\mu x^\mu} + \ud{\hat{b}}_{\, p}^{\wedge/\vee \, \da} v_p^{\wedge/\vee} e^{+i p_\mu x^\mu} \rp^* \\[5pt]
&=& \int{\fr{d^3p}{(2\pi)^3 (2E)}} \lp  \ud{\hat{a}}_{\, p}^{\wedge/\vee \, \da} v_p^{\wedge/\vee} e^{+i p_\mu x^\mu} + \ud{\hat{b}}_{\, p}^{\wedge/\vee} u_p^{\wedge/\vee} e^{-i p_\mu x^\mu} \rp
\ea
$$
using Dirac solution identities, $i \ga_2  \, u_p^{\wedge/\vee \, *}  =  v_p^{\wedge/\vee}$ and $i \ga_2  \, v_p^{\wedge/\vee \, *}  =  u_p^{\wedge/\vee}$. The charge conjugation transformations of the creation and annihilation operators, using the corresponding unitary operation on the infinite-dimensional representation, $\ud{\hat{\Ps}}^C = \hat{C} \ud{\hat{\Ps}} \hat{C}^-$, are thus:
\beq
\lp \ud{\hat{a}}_{\, p}^{\wedge/\vee} \rp^C =  \ud{\hat{b}}_{\, p}^{\wedge/\vee}
\s \s
\lp \ud{\hat{b}}_{\, p}^{\wedge/\vee \, \da} \rp^C =  \ud{\hat{a}}_{\, p}^{\wedge/\vee \, \da}
\label{charge}
\eeq

The unitary parity conjugate, $\Ps^P = i \ga_0 \Ps(t,-x)$, is 
$$
\ba{rcl}
\ud{\hat{\Ps}}^P &=& i \ga_0 \int{\fr{-d^3p}{(2\pi)^3 (2E)}} \lp \ud{\hat{a}}_{\, -p}^{\wedge/\vee} u_{-p}^{\wedge/\vee} e^{-i p_\mu x^\mu} + \ud{\hat{b}}_{\, -p}^{\wedge/\vee \, \da} v_{-p}^{\wedge/\vee} e^{+i p_\mu x^\mu} \rp \\[5pt]
&=& \int{\fr{d^3p}{(2\pi)^3 (2E)}} \lp -i \, \ud{\hat{a}}_{\, -p}^{\wedge/\vee} u_p^{\wedge/\vee} e^{-i p_\mu x^\mu} +i \, \ud{\hat{b}}_{\, -p}^{\wedge/\vee \, \da} v_p^{\wedge/\vee} e^{+i p_\mu x^\mu} \rp
\ea
$$
using $i \ga_0 \, u_{-p}^{\wedge/\vee} =  + i \, u_p^{\wedge/\vee}$ and $i \ga_0 \, v_{-p}^{\wedge/\vee} = - i \, v_p^{\wedge/\vee}$. The unitary parity conjugation transformations of the creation and annihilation operators are thus:
\beq
\lp \ud{\hat{a}}_{\, p}^{\wedge/\vee} \rp^P = -i \, \ud{\hat{a}}_{\, -p}^{\wedge/\vee}
\s \s
\lp \ud{\hat{b}}_{\, p}^{\wedge/\vee \, \da} \rp^P = +i \, \ud{\hat{b}}_{\, -p}^{\wedge/\vee \, \da}
\label{parity}
\eeq

The unitary time conjugate, $\Ps^T  = \ga_{13} \Ps(-t,x)$, of a massless quantum Dirac spinor corresponds to an antiunitary operator on Fock space,
$$
\ba{rcl}
\ud{\hat{\Ps}}^T &=& \hat{T}' \ud{\hat{\Ps}} \hat{T}'^- 
= \int{\fr{d^3p}{(2\pi)^3 (2E)}} \lp \lp \ud{\hat{a}}_{\, p}^{\wedge/\vee} \rp^T u_p^{\wedge/\vee \, *} e^{+i p_\mu x^\mu} + \lp \ud{\hat{b}}_{\, p}^{\wedge/\vee \, \da} \rp^T v_p^{\wedge/\vee \, *} e^{-i p_\mu x^\mu} \rp \\[5pt] 
&=& \ga_{13} \int{\fr{d^3p}{(2\pi)^3 (2E)}} \lp \ud{\hat{a}}_{\, -p}^{\wedge/\vee} u_{-p}^{\wedge/\vee} e^{+i p_\mu x^\mu} + \ud{\hat{b}}_{\, -p}^{\wedge/\vee \, \da} v_{-p}^{\wedge/\vee} e^{-i p_\mu x^\mu} \rp \\[5pt]
&=&  \int{\fr{-d^3p}{(2\pi)^3 (2E)}} \lp \mp \ud{\hat{a}}_{\, -p}^{\wedge/\vee} u_p^{\vee/\wedge \, *} e^{+i p_\mu x^\mu} \mp \ud{\hat{b}}_{\, -p}^{\wedge/\vee \, \da} v_p^{\vee/\wedge \, *} e^{-i p_\mu x^\mu} \rp \\
\ea
$$
using $\ga_{13}  \, u_{-p}^{\wedge/\vee}  = \mp \,  u_p^{\vee/\wedge \, *}$ and $\ga_{13} \, v_{-p}^{\wedge/\vee}  = \mp  \,  v_p^{\vee/\wedge \, *}$. The time conjugation transformations of the creation and annihilation operators for particles and antiparticles are thus:
\beq
\lp \ud{\hat{a}}_{\, p}^{\wedge/\vee} \rp^T = \mp \, \ud{\hat{a}}_{\, -p}^{\vee/\wedge}
 \s \s
\lp \ud{\hat{b}}_{\, p}^{\wedge/\vee \, \da} \rp^T = \mp \, \ud{\hat{b}}_{\, -p}^{\vee/\wedge \, \da}
\label{time}
\eeq
The existence of this antiunitary time conjugation operator, $T$, squaring to minus one, implies our fermion representation space is quaternionic.\cite{Baez}

Applied to the weight vectors of a massless quantum Dirac spinor along the $z$ axis, these conjugations give maps between particle states and between their weights:
\beq
\ba{rclcrcl}
(a_{L/R}^{\wedge/\vee})^C &=& \bar{a}_{L/R}^{\wedge/\vee} & \s \s & C \; : \;  (\om_T, \om_S, h, q) &\; \mapsto \;& (\;\;\, \om_T \;\;\, \om_S, \;\;\, h, - q) \\[10pt]
(a_{L/R}^{\wedge/\vee})^P &=& -i \, a_{R/L}^{\wedge/\vee} &  & P \; : \; (\om_T, \om_S, h, q) &\; \mapsto \;& (-\om_T, \;\;\, \om_S, - h, \;\;\, q) \\[10pt]
(a_{L/R}^{\wedge/\vee})^T &=& \mp \, a_{L/R}^{\vee/\wedge} &  & T \; : \; (\om_T, \om_S, h, q) &\; \mapsto \;& (-\om_T, - \om_S, \;\;\, h, \;\;\, q) \\
\ea
\label{CPTweights}
\eeq
Plotting fermion weights, $(\om_T, \om_S, q)$, and their conjugation relationships, we get the CPT cube:
\begin{center}
\begin{tikzpicture}[scale=1.7]
  \coordinate (A) at (1,1,1);
  \coordinate (B) at (1,1,-1);
  \coordinate (C) at (1,-1,1);
  \coordinate (D) at (-1,1,1);
  \coordinate (E) at (1,-1,-1);
  \coordinate (F) at (-1,1,-1);
  \coordinate (G) at (-1,-1,1);
  \coordinate (H) at (-1,-1,-1);
  \draw (A) -- (C) node[midway, right=-2pt] {$C$};
  \draw (C) -- (G) node[midway, below=-2pt] {$P$};
  \draw (G) -- (D) node[midway, left=-2pt] {$C$};
  \draw (D) -- (A) node[midway, above=-2pt] {$P$};
  \draw (B) -- (E) node[midway, right=-2pt] {$C$};
  \draw (E) -- (H) node[midway, below=-2pt] {$P$};
  \draw (H) -- (F) node[midway, left=-2pt] {$C$};
  \draw (F) -- (B) node[midway, above=-2pt] {$P$};
  \draw (F) -- (E) node[midway, left=3pt] {$CP$};
  \draw (D) -- (C) node[midway, above=6pt] {$CP$};
  \draw (A) -- (B) node[midway, above =3pt] {$PT$};
  \draw (C) -- (E) node[midway, below =4pt] {$PT$};
  \draw (G) -- (H) node[midway, below =4pt] {$PT$};
  \draw (D) -- (F) node[midway, above =3pt] {$PT$};
  \draw (F) -- (A) node[midway, above right=-3pt] {$T$};
  \draw (C) -- (H) node[midway, below left=-3pt] {$T$};
  \draw (A) -- (H) node[midway, below right=-3pt] {$CT$};
  \draw (B) -- (C) node[midway, above = 7pt] {$CPT$};
  \draw (F) -- (G) node[midway, below = 6pt] {$CPT$};
  \node[scale=1.5] at (B) {\gpru{myell}};
  \node[scale=1.5] at (A) {\gpld{myell}};
  \node[scale=1.5] at (D) {\gprd{myell}};
  \node[scale=1.5] at (F) {\gplu{myell}};
  \node[scale=1.5] at (C) {\gald{myell}};
  \node[scale=1.5] at (E) {\garu{myell}};
  \node[scale=1.5] at (G) {\gard{myell}};
  \node[scale=1.5] at (H) {\galu{myell}};
  \node[above=3pt] at (B) {$f_R^\wedge$};
  \node[above=3pt] at (A) {$f_L^\vee$};
  \node[above=3pt] at (D) {$f_R^\vee$};
  \node[above=3pt] at (F) {$f_L^\wedge$};
  \node[below=3pt] at (C) {$\bar{f}_L^\vee$};
  \node[below=3pt] at (E) {$\bar{f}_R^\wedge$};
  \node[below=3pt] at (G) {$\bar{f}_R^\vee$};
  \node[below=3pt] at (H) {$\bar{f}_L^\wedge$};
\end{tikzpicture}
\end{center}

\noindent Composition of the $C$, $P$, and $T$ operators produces the CPT Group, $G_{CPT } = Q_8 \times \mathbb{Z}_2$, of order $16$, equivalent to the split-biquaternion group. To understand this equivalence, we can identify the charge conjugation operator, $C \sim I$, with a split-complex number, $I^2=1$, which commutes with parity and time conjugation operators identified with unit quaternions, $P \sim e_3$, $T \sim e_2$, and $PT \sim e_3 e_2 = - e_1$. The $G_{CPT}$ multiplication table is:

\begin{table}[h!t]
\centering
\begin{tabular}
{@{\vrule width1.0pt}c@{\vrule width0.4pt}c@{\vrule width0.4pt}c@{\vrule width0.4pt}c@{\vrule width0.4pt}c@{\vrule width0.4pt}c@{\vrule width0.4pt}c@{\vrule width0.4pt}c@{\vrule width1.0pt}}
 \noalign{\hrule height 1.0pt}
  $1$ & $C$ & $P$ & $T$ & $CP$ & $CT$ & $PT$ & $\; CPT \;$ \\[-1pt]
 \noalign{\hrule height 0.2pt}
  $C$ & $+1$ & $CP$ & $CT$ & $P$ & $T$ & $\; CPT \;$ & $PT$ \\[-1pt]
 \noalign{\hrule height 0.2pt}
  $P$ & $CP $ & $-1 $ & $PT $ & $-C $ & $\; CPT \; $ & $-T $ & $-CT $ \\[-1pt]
 \noalign{\hrule height 0.2pt}
  $ T$ & $CT $ & $-PT $ & $-1 $ & $\,-CPT\,$ & $-C $ & $P $ & $CP $ \\[-1pt]
 \noalign{\hrule height 0.2pt}
  $ CP$ & $P $ & $-C $ & $\;CPT\; $ & $-1 $ & $PT $ & $-CT $ & $-T $ \\[-1pt]
   \noalign{\hrule height 0.2pt}
  $ CT$ & $T $ & $\,-CPT\,$ & $-C $ & $-PT $ & $-1 $ & $CP $ & $P $ \\[-1pt]
 \noalign{\hrule height 0.2pt}
  $ PT$ & $\;CPT \;$ & $T $ & $-P $ & $CT $ & $-CP $ & $-1 $ & $-C $ \\[-1pt]
   \noalign{\hrule height 0.2pt}
  $\; CPT \;$ & $PT $ & $CT $ & $-CP $ & $T $ & $-P $ & $-C $ & $-1 $ \\[-1pt]
  \noalign{\hrule height 1.0pt}
\end{tabular}
\caption{The CPT Group multiplication table---with further multiplications by $-1$ implied.}
\label{table:CPT}
\end{table}

\newpage

\section{Biquaternionic Spinors, Triality, and the CPTt Group}

Because fermions exist in three generations, we can introduce a natural fourth discrete conjugation operator, triality ($t$), that maps between generations and satisfies $t^3 = 1$. One nontrivial extension of the $CPT$ Group, $G_{CPT} = Q_8 \times \mathbb{Z}_2$, to a group, $G_{CPTt'}$, acting on three generations of fermions, can be constructed by identifying a triality generator element, such as $t' \sim - \ha (1 + e_1 + e_2 + e_3)$.\cite{Wilson} This choice of triality generator, $t'$, commutes with the split-complex generator, $C \sim I$, and the adjoint, $\text{ad}_{t'}$, cycles imaginary quaternion basis elements, such as $\text{ad}_{t'} e_1 = t' e_1 t'{}^2 = e_2$. This triality generator (either $\text{ad}_{t'}$ or $t'$) extends the $PT$ group, $G_{PT}=Q_8$, to the binary-tetrahedral group, $G_{PTt'} = 2T$, and including charge conjugation via the split-complex generator gives the $CPTt'$ Group, $G_{CPTt'} = 2T \times \mathbb{Z}_2$, the split-binary-tetrahedral group, of order $48$. Although this is mathematically sound, this is not the only choice of group extension by triality. The Standard Model is not invariant under charge conjugation, $C$, so it is not expected that our $t$ symmetry should commute with $C$. It is the case that the Standard Model is invariant under $CPT$, so what we need is to have different $C$, $P$, and $T$ representatives that don't commute with our $t$, such that the resulting $CPT$ generator does commute with $t$. Rather than guess at such new group representatives, we can revisit the operation of $C$, $P$, and $T$ on Dirac spinors, and translate these to operations on biquaternionic spinors.

A Dirac spinor describes both a fermion and an antifermion, via positive and negative energy Dirac solutions. This suggests a re-arrangement of degrees of freedom, using the charge conjugate,
$$
\Ps = \lb \ba{c} \Ps_1 \\ \Ps_2 \\ \Ps_3 \\ \Ps_4 \ea \rb \s
\Ps^C = i \ga_2 \Ps^* = \lb \ba{c} -\Ps_4^* \\ \Ps_3^* \\ \Ps_2^* \\ -\Ps_1^* \ea \rb \s
\Ps_Q = \lb \ba{cc} \Ps_1 & -\Ps_4^* \\ \Ps_2 & \Ps_3^* \\ \Ps_3 & \Ps_2^* \\ \Ps_4 & -\Ps_1^* \ea \rb
\sim \lb \ba{c} \ps_{\mathbb{H}L} \\ \ps_{\mathbb{H}R} \ea \rb
$$
in which all Dirac spinor degrees of freedom can inhabit either the left or right-chiral \emph{Dirac biquaternions}, $\ps_{\mathbb{H}L}$ or  $\ps_{\mathbb{H}R}$. Here we make use of the representation of quaternions, $e_\mu \in \mathbb{H}$, using Pauli matrices, $\{e_0 \!\sim\! \si_0, \, e_\pi \!\sim\! - i \si_\pi \}$, and the definition of biquaternions as quaternions with complex coefficients, $ \ps_{\mathbb{H}L} \in \mathbb{C} \otimes \mathbb{H}$.
$$
\ps_{\mathbb{H}L} = \ps^\mu_{\mathbb{H}L} e_\mu
\sim \ps_{QL} = 
\lb \ba{cc}
\Ps_1^\mathbb{R} + i \Ps_1^\mathbb{I} &\, - \Ps_4^\mathbb{R} + i \Ps_4^\mathbb{I} \\
\Ps_2^\mathbb{R} + i \Ps_2^\mathbb{I} &\, \Ps_3^\mathbb{R} - i \Ps_3^\mathbb{I}
 \ea \rb
= \lb \ps_L \; \bar{\ps}_L \rb
 \q
\ps_{QR} = i \si_2 \ps_{QL}^* \si_1 \sim  \ps_{\mathbb{H}R} = i \ps_{\mathbb{H}L}^* e_3
$$
$$
\ba{rcl}
\ps_{\mathbb{H}L} &=& \ha \lp \Ps_1^\mathbb{R} + \Ps_3^\mathbb{R} - i \Ps_1^\mathbb{I} + i \Ps_3^\mathbb{I} \rp e_0
+ \ha \lp \Ps_2^\mathbb{I} + \Ps_4^\mathbb{I} + i \Ps_2^\mathbb{R} - i \Ps_4^\mathbb{R} \rp e_1 \\[2pt]
& + & \ha \lp -\Ps_2^\mathbb{R} - \Ps_4^\mathbb{R} + i \Ps_2^\mathbb{I} - i \Ps_4^\mathbb{I} \rp e_2
+ \ha \lp \Ps_1^\mathbb{I} + \Ps_3^\mathbb{I} + i \Ps_1^\mathbb{R} - i \Ps_3^\mathbb{R} \rp e_3
\ea
$$
Describing the biquaternions and their representation requires juggling several conjugations. Since the Pauli matrices satisfy $\si_\mu^* = \si_2 \bar{\si}_\mu \si_2$, we can define a similar conjugation for biquaternions, $\ps_{QL}^* \sim - e_2 \ps_{\mathbb{H}L}^* e_2$; and since the Pauli matrices are Hermitian, we also have $\ps_{QL}^\da \sim \tilde{\ps}{}_{\mathbb{H}L}^*$, using complex and quaternionic conjugation, $\{ \tilde{e}_0 = e_0, \tilde{e}_\pi = - e_\pi \}$. The invariant bilinear form on the biquaternions directly relates to the bilinear Dirac spinor scalar, 
$$
\ba{rcl}
( \ps_{\mathbb{H}L},\ps_{\mathbb{H}L} )  &=& \lp \Ps_1^\mathbb{R} \Ps_3^\mathbb{R} + \Ps_2^\mathbb{R} \Ps_4^\mathbb{R} +\Ps_1^\mathbb{I} \Ps_3^\mathbb{I} + \Ps_2^\mathbb{I} \Ps_4^\mathbb{I} \rp
+ i \lp - \Ps_1^\mathbb{R} \Ps_3^\mathbb{I} - \Ps_2^\mathbb{R} \Ps_4^\mathbb{I} +\Ps_1^\mathbb{I} \Ps_3^\mathbb{R} + \Ps_2^\mathbb{I} \Ps_4^\mathbb{R} \rp \\[4pt]
&=& \tilde{\ps}_{\mathbb{H}L} \ps_{\mathbb{H}L} = \det(\ps_{QL}) \\[4pt]
\bar{\Ps} \Ps &=& 2 \lp \Ps_1^\mathbb{R} \Ps_3^\mathbb{R} + \Ps_2^\mathbb{R} \Ps_4^\mathbb{R} +\Ps_1^\mathbb{I} \Ps_3^\mathbb{I} + \Ps_2^\mathbb{I} \Ps_4^\mathbb{I} \rp
= 2 \, \Re(\tilde{\ps}_{\mathbb{H}L} \ps_{\mathbb{H}L})
\ea
$$
We can now calculate $C$, $P$, and $T$ symmetry conjugations for our biquaternionic fermions,
\beq
\ba{rclcrclcrclcrcl}
\Ps^C &=& i \ga_2 \Ps^* &\s& \ps_{QL}^C &=& \ps_{QL} \si_1 &\s& \ps_{\mathbb{H}L}^C &=& i \ps_{\mathbb{H}L} e_1 &\s& C &\sim& i e_1 \\[2pt]
\Ps^P &=& i \ga_0 \Ps &\s& \ps_{QL}^P &=& - \si_2 \ps_{QL}^* \si_1 &\s& \ps_{\mathbb{H}L}^P &=& -  \ps_{\mathbb{H}L}^* e_3 &\s& P &\sim& -K e_3 \\[2pt]
\Ps^{T} &=& \ga_{13} \Ps^* &\s& \ps_{QL}^{T} &=& i \si_2 \ps_{QL}^* &\s& \ps_{\mathbb{H}L}^{T} &=& - \ps_{\mathbb{H}L}^* e_2 &\s& T &\sim& -K e_2
\ea
\label{biquatCPT}
\eeq
in which we use the antiunitary time conjugation operator, $U_T$, and correctly reproduce the $CPT$ Group action on fermions, with quaternionic multiplication from the right. From these $C$, $P$, and $T$ generators we have $CPT \sim -i$.

We can now add a triality generator, $t \sim - \ha (1 + e_1 + e_2 + e_3)$ or $\text{ad}_t$, to those of our new $C$, $P$, and $T$ representatives and see that this does not commute with $C$ and does commute with $CPT$. With the addition of $t$, since we have $P T \sim e_2 e_3 = e_1$, and $\text{ad}_t$ cycles imaginary quaternions, we can also construct an expression for the complex conjugation generator in terms of other generators, $K \sim t P T t t T$. This $K$ generator commutes with $P$, $T$, and $t$, but anticommutes with $C$. It is antiunitary, corresponding to complex conjugation of a Dirac spinor, $\Ps^K = K \Ps = \Ps^*$, and corresponds to creation conjugation on the infinite-dimensional representation space generators of QFT,
$$
\begin{array}{rcl}
\ud{\hat{\Ps}}^K &=& \hat{K} \ud{\hat{\Ps}} \hat{K}^- 
= \int{\fr{d^3p}{(2\pi)^3 (2E)}} \lp \lp \ud{\hat{a}}_{p}^{\wedge/\vee} \rp^K {u}_{p}^{\wedge/\vee \, *} e^{+i p_\mu x^\mu} + \lp \ud{\hat{b}}_{p}^{\wedge/\vee \, \da} \rp^K {v}_{p}^{\wedge/\vee \, *} e^{- i p_\mu x^\mu} \rp \\[6pt] 
&=& \int{\fr{d^3p}{(2\pi)^3 (2E)}} \lp \ud{\hat{a}}_p^{\wedge/\vee} {u}_p^{\wedge/\vee} e^{-i p_\mu x^\mu} + \ud{\hat{b}}_p^{\wedge/\vee \, \da} {v}_p^{\wedge/\vee} e^{+i p_\mu x^\mu} \rp^* \\[6pt]
&=& \int{\fr{d^3p}{(2\pi)^3 (2E)}} \lp \ud{\hat{a}}_p^{\wedge/\vee \, \da} {u}_p^{\wedge/\vee \, *} e^{+i p_\mu x^\mu} + \ud{\hat{b}}_p^{\wedge/\vee} {v}_p^{\wedge/\vee \, *} e^{-i p_\mu x^\mu} \rp \\
\end{array}
$$
with
$$
\lp \ud{\hat{a}}_p^{\wedge/\vee} \rp^K = \ud{\hat{a}}_{p}^{\wedge/\vee \, \da} \s\;
\lp \ud{\hat{b}}_p^{\wedge/\vee} \rp^K = \ud{\hat{b}}_{p}^{\wedge/\vee \, \da}
$$
Creation conjugation changes particle annihilation into particle creation, mapping positive energy states to nonphysical negative energy states. This is not considered a symmetry of nature---but it is part of our symmetry group.

The $CPTt$ Group is generated by:
\beq
C \sim i e_1 \s \s P \sim -K e_3 \s \s T \sim -K e_2 \s \s t \sim -\ha (1 + e_1 + e_2 + e_3) \;\; \text{or} \;\; \text{ad}_t
\label{fungen}
\eeq
The $P$ and $T$ generators produce a quaternion subgroup, $Q_8$, and the triality generator, $t$ or $\text{ad}_t$, extends this to the binary tetrahedral group, $2T$, which includes $K$. Including $C$ produces $CPT \sim -i$ and extends $2T$ to the $CPTt$ Group, $G_{CPTt}$, of order $96$. The $CPTt$ Group is the central product, $G_{CPTt} = 2T \circ D_4$, of $2T$ and $D_4$, the dihedral group of order $8$. It has $Q_8 = \{ \pm 1, \pm e_1, \pm e_2, \pm e_3 \}$, $2T$, and $D_4 = \{ \pm 1, \pm i, \pm K, \pm i K \}$, as normal subgroups, with a shared center, $\mathbb{Z}_2 = \{1,-1\}$. Consulting GAP, of $231$ finite groups of order $96$, only one has $2T$ and $D_4$ normal subgroups with a shared central $\mathbb{Z}_2$---this is the $CPTt$ Group, $G_{CPTt} = 2T \circ D_4$ (GAP ID $[96, 202]$).{\cite{Gap}}

Just as the $CPT$ Group acts on $8$ quantized Dirac fermion states projectively represented as the $CPT$ cube in three dimensions, with vertices corresponding to fermion weights, $(\om_T, \om_S, q)$, the $CPTt$ Group acts on three generations of fermion states projectively represented as a 24-cell, which lives naturally in four dimensions. For the four weight coordinates of the 24-cell we can choose $(\om_t, \om_S, h, q)$, in which $\om_t = 4 \om_S h q$ is a \emph{Euclidean boost weight}, such that the eight weights of one fermion correspond to the weights of an octonion under $spin(8)$. In these coordinates the projective group action changes the signs of the first generation fermions by matrices, $C_I$, $P_I$, and $T_I$, and the second and third generation particles have weights related to the first via multiplication times a \emph{triality matrix}, $t$. The resulting $CPTt$ Group projective representation generators are:
$$
C_I = \lb \ba{cccc}
- & & &  \\[-4pt]
 & + &  &  \\[-4pt]
 &  & + &  \\[-4pt]
 &  &  & - \\
\ea  \rb
\s
P_I = \lb \ba{cccc}
- & & &  \\[-4pt]
 & + &  &  \\[-4pt]
 &  & - &  \\[-4pt]
 &  &  & + \\
\ea  \rb
\s
T_I = \lb \ba{cccc}
- & & &  \\[-4pt]
 & - &  &  \\[-4pt]
 &  & + &  \\[-4pt]
 &  &  & + \\
\ea  \rb
\s
t = \ha \! \lb \ba{cccc}
+ & - & + & + \\[-4pt]
+ & - & - & - \\[-4pt]
+ & + & + & - \\[-4pt]
+ & + & - & + \\
\ea  \rb
$$
The weights of the second and third generation particles are necessarily nonsensical under direct interpretation, but are correct---matching those of the first generation---when considered under triality. Similarly, the action of $C$, $P$, and $T$ on the second (and similarly on the third) generation particles are by the matrices $C_{II} = t C_I t^2$, $P_{II} = t P_I t^2$, and $T_{II} = t T_I t^2$. To better understand what triality is, and where this triality matrix comes from, we need to understand division algebras.{\cite{Div}} 

\begin{figure}[h!t]
\begin{center}
\includegraphics[height=2.3in]{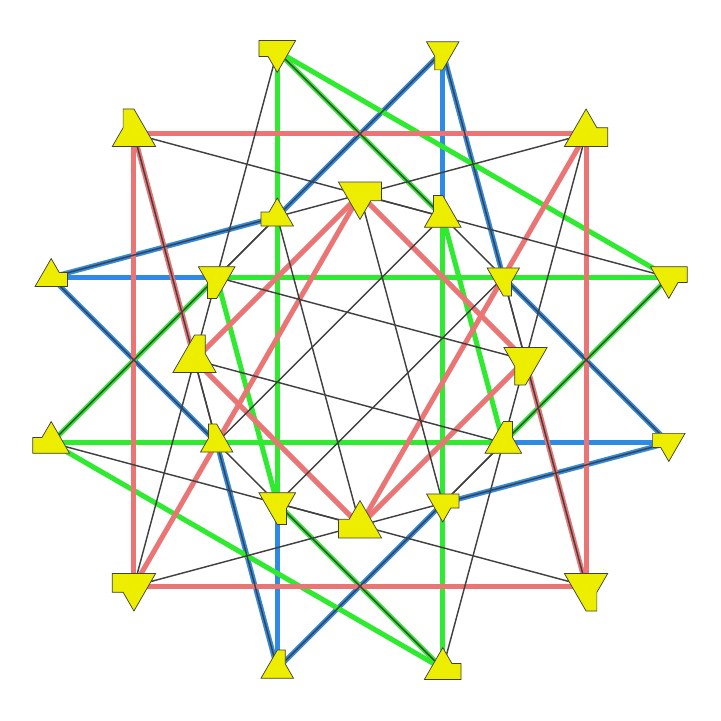} \\
\caption{The $24$ elementary particle states of three generations of massless quantum Dirac fermion states (such as the electron, muon, and tau) represented as a 24-cell, acted on by the $CPTt$ Group. The $CPT$ cube of the $8$ first generation states is shown with red edges, the second generation $CPT$ cube in green, and third generation $CPT$ cube in blue. The three generations are linked by triality, $t$, shown in black, with second and third generation fermion states shown with smaller glyphs.}
\end{center}
\end{figure}
 
\newpage 
 
\section{Discussion} 
 
The fundamental symmetries of a quantum Dirac spinor are the charge (\ref{charge}), parity (\ref{parity}), and time (\ref{time}) conjugations of the corresponding creation and annihilation operators. Charge conjugation swaps particles and antiparticles, parity conjugation reflects momentum and swaps left and right helicities (preserving spin), and time conjugation reflects momentum and swaps spin (preserving helicity). These conjugations correspond directly to operations on Dirac spinors or the corresponding biquaternionic spinors, and to projective actions on their weights. Time conjugation must be handled especially carefully, via a unitary $U^Q_T = \ga_{13}$ on Dirac spinors, or an antiunitary $U_T = \ga_{13} K$ operator that preserves the group structure. The finite group generated by these $C$, $P$, and $T$ conjugations is the $CPT$ Group, $G_{CPT} = Q_8 \times \mathbb{Z}_2$, equivalent to the split-biquaternion group. This group acts projectively on the $8$ weights of fermion states in the $CPT$ cube.
 
Using an isomorphism to biquaternionic Dirac spinors, the $C = i e_1$, $P = - K e_3$, and $T=-K e_2$ generators (\ref{biquatCPT}) of the $CPT$ Group are extended by a quaternionic triality generator, $t = \ha(1+e_1+e_2+e_3)$ or the adjoint, $\text{ad}_t$, commuting with the $CPT$ generator, $CPT = -i$, to produce the $CPTt$ Group, $G_{CPTt} = 2T \circ D_4$, of order 96, a central product of $2T$ and $D_4$. This group acts projectively on the $24$ weights (the vertices of a 24-cell) corresponding to three generations of each fermion type and its corresponding three $CPT$ cubes. It is important to note that the weights (spins and charges) of the second and third generation fermions described this way are nonsensical when considered as weights of the Lorentz algebra acting on the first generation---they only make sense as weights related to first-generation charges by triality.

The identification of triality, $t$, as a partner to $C$, $P$, and $T$ symmetries, and the extension to the $CPTt$ Group, seems likely to be of fundamental importance in the Standard Model and its unification with gravity. Although it is possible to trivially extend the $CPT$ Group to a direct product $CPTt$ Group, such as extending the $CPT$ Group to $G = G_{CPT} \times A_4$, such extension seems unlikely to present the rich and varied generational mixing in the Standard Model; we prefer a more unified model, with a non-direct-product $CPTt$ Group. In the complete Standard Model there are $8$ fermion types: neutrinos, electron-type leptons, three colors of up-type quarks, and three colors of down-type quarks. In a unified theory (which includes right-handed neutrinos) these must correspond to $8$ disjoint 24-cells, each transforming under the $CPTt$ group. The only current proposal for a unified theory that meets this criteria, with a triality symmetry acting between $192$ distinct fermion weights grouped into $8$ disjoint 24-cells, is E8 Theory.{\cite{Lisi1,Lisi2}}

\newpage
 
\begin{figure}[h!t]
\begin{center}
\includegraphics[height=6.0in]{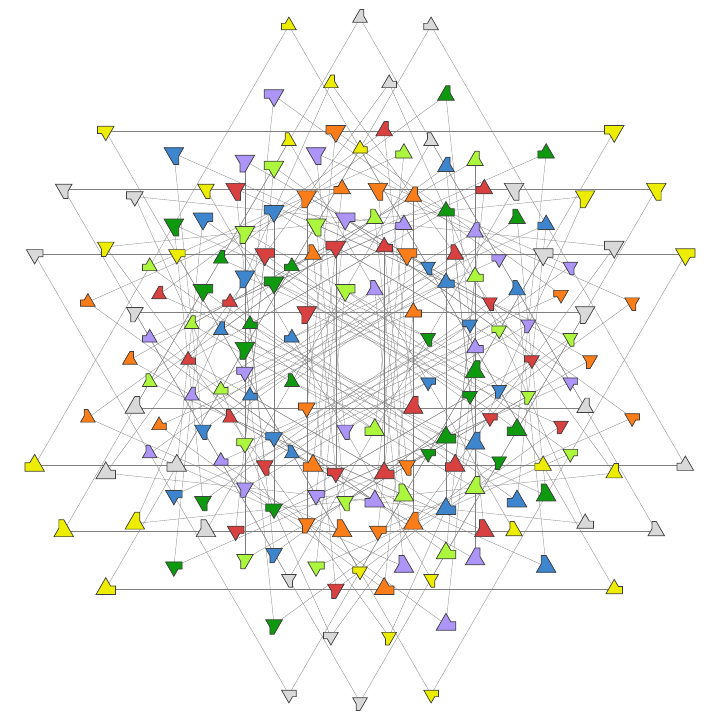} \\
\caption{The $192$ elementary particle states of three generations of massless quantum Dirac fermion states, including $8$ disjoint 24-cells, with each 24-cell corresponding to a different type of fundamental fermion: neutrinos (gray), electron-type leptons (yellow), up-type quarks (red, green, and blue) and down-type quarks (orange, chartreuse, and purple). Each 24-cell includes a triality-related triplet (connected by gray lines, with second and third generation fermions shown with smaller glyphs) of $8$ fermion states (cubes, not shown) related by $C$, $P$, and $T$ conjugations.}
\end{center}
\label{fig:CPTt192}
\end{figure}


\newpage
 
\begin{thebibliography}{99}

\bibitem{Pin} M. Berg, C. DeWitt-Morette, S. Gwo, E. Kramer, ``The Pin Groups in Physics: C, P, and T,'' \arXiv{math-ph/0012006} 

\bibitem{CPT} M. Socolovsky, ``The CPT Group of the Dirac Field,'' \arXiv{math-ph/0404038} 

\bibitem{Wilson} R. A. Wilson, ``Finite Symmetry Groups in Physics,'' \arXiv{2102.02817}

\bibitem{Baez} J. Baez, ``Division Algebras and Quantum Theory,'' \arXiv{1101.5690}

\bibitem{Gap} The GAP group (2022), \emph{GAP – Groups, Algorithms, and Programming}, \textbf{4.12.1}, \href{https://www.gap-system.org}{https://www.gap-system.org}

\bibitem{Div} A. G. Lisi, ``Division Algebras, Triality, and Exceptional Magic,'' in preparation

\bibitem{Lisi1} A. G. Lisi, ``An Exceptionally Simple Theory of Everything,'' \arXiv{0711.0770}

\bibitem{Lisi2} A. G. Lisi, ``Exceptional Unification,'' in preparation

\end{thebibliography}
\end{document}